\documentclass[prb,superscriptaddress,longbibliography,notitlepage,showpacs,twocolumn]{revtex4-2}
\usepackage{tikz}
\usepackage{color}
\usepackage{float,xcolor}
\usepackage{amsmath}
\usepackage{amssymb}
\usepackage{graphicx,array}
\usepackage{txfonts}
\usepackage{mathtools}
\usepackage{xspace}
\usepackage{ulem}
\newcommand{\Tr}{{\rm Tr}}

\newcommand{\dg}{\dagger}

\newcommand{\calh}{\mathcal{H}}
\newcommand{\beq}{\begin{equation}}
\newcommand{\eeq}{\end{equation}}

\newcommand{\rhot}{\tilde\rho}
\graphicspath{{Figures/}}

\begin{document}

\title{Mpemba effects in nonequilibrium open quantum systems}

\author{Xuanhua Wang}
\email{wangxh@ucas.ac.cn}
\affiliation{Center for Theoretical Interdisciplinary Sciences, Wenzhou Institute, University of Chinese Academy of Sciences, Wenzhou, Zhejiang 325001, China}

\author{Jin Wang}
\email{Corresponding author: jin.wang.1@stonybrook.edu}
\affiliation{Department of Physics and Astronomy, Stony Brook University, Stony Brook, New York 11794, USA}
\affiliation{Department of Chemistry, Stony Brook University, Stony Brook, New York 11794, USA}


\begin{abstract}
The Mpemba effect was originally referred to as the faster icing of a higher-temperature system than a lower-temperature system, and was later generalized to anomalous decays of both classical and quantum observables to equilibrium states. Mpemba effect is mostly considered in classical systems and during cooling processes towards equilibrium states. We investigate the emergence of the effect in nonequilibrium quantum systems where the system has no asymptotic equilibrium state to approach. Instead, the system is put in contact with two different baths, and only a nonequilibrium state exists, sustained by constant energy injection from the surrounding thermal baths. Firstly, we show that the nonequilibrium conditions can dramatically enlarge the parameter regimes where the MPE emerges. Secondly, we demonstrate that the anomalous MPEs and inverse MPEs emerge in the evolution of quantum correlations in the two-site fermionic system and that nonequilibrium conditions can expedite or delay the MPEs. Thirdly, we show that the nonequilibrium-induced quantum coherence can have considerable contributions to the emergence of the MPE which the conventional Lindbladian dynamics fails to capture.
\end{abstract}
\maketitle
%
%
\thispagestyle{empty}


\section*{Introduction}
The anomalous cooling process in which hotter liquids freeze faster than colder liquid has been documented long before the inception of the modern science. The first reported experimental study of the phenomenon in classical systems was conducted in 1960s by Mpemba and Osborne \cite{mpemba1969cool}, and has triggered heated debates \cite{bechhoefer2021fresh,burridge2016questioning}. Mpemba effect (MPE) has been observed in multiple substances by now such as water \cite{jeng2006mpemba}, nanotube resonators \cite{greaney2011mpemba}, granular fluids \cite{lasanta2017hotter,biswas2020mpemba}, and Langevin systems \cite{biswas2023mpemba}. Consequently, the concept of MPE has been generalized from its original context to a broader class of anomalous decays such as the exponentially-accelerated relaxations and the anomalous crossings of many system observables \cite{lu2017nonequilibrium,klich2019mpemba}. Recently, seminal works on models such as quantum dots and spin chains have revealed MPEs in the quantum realm \cite{carollo2021exponentially,qdot2023,chatterjee2023multiple,ares2023entanglement,bao2022accelerating,murciano2024entanglement,rylands2023microscopic}, and quantum Mpemba effects (QMPEs) have been experimentally observed in ion traps \cite{shapira2024mpemba,joshi2024observing}. In a classical quasi-static cooling process, the trajectories of a hotter and a cooler systems never intercept according to Newton's heat law; similarly, in a closed quantum system, the distinctions of states are protected by the unitarity and different states do not intercept in the evolution--a principle known as the conservation of information. Due to the violation of the above principles, it is believed that MPE emerges only in open and strongly nonequilibrium systems \cite{lu2017nonequilibrium,carollo2021exponentially}. 

Though multiple mechanisms for MPEs have been hypothesized, whether a universal mechanism for MPEs in different systems exists is not completely clear \cite{esposito2008mpemba,vynnycky2010evaporative,zhang2014hydrogen,jin2015mechanisms}. The first widely applicable mechanism in the Markovian environment was proposed by analyzing the Markovian dynamics of a classical three-level system and the Ising model \cite{lu2017nonequilibrium}. The evolution of a system in a Markovian process follows the linear equation
\begin{gather}
    \frac{d\Vec{p}(t)}{dt}=\mathcal{L}_{T_b} \Vec{p}(t)
\end{gather}
where $\mathcal{L}_{T_b}$ is the linear Markovian operator and $T_b$ is the bath temperature. For a N-level classical system, the Markovian operator can be written using the transition rate matrix $R_{ij}$, i.e., $\frac{dp_i(t)}{dt}=\sum_{j} R_{ij}p_j(t)$. This linear equation can be formally solved by 
\begin{gather}
    \Vec{p}(t)=e^{\hat R t}\Vec{p}(0)=\sum_i e^{\lambda_i t} \alpha_i \Vec{v}_i\,,
\end{gather}
where $\Vec{v}_i$ is the $i$-th right eigenvector of the transition matrix $\hat R$ and $\alpha_i$ is the coefficient of $\Vec{p}(0)$ projecting onto the left eigenvector associated with $\Vec{v}_i$. The core of the argument is that for certain initial states and bath temperatures $T_b$, the distribution vector $\Vec{p}(0)$ has tiny or zero component on the eigenvector of the slowest decaying mode. This will result in the exponentially faster relaxation of the system compared to a randomly-picked state. Notably, this analysis can be easily generalized into the quantum realm by replacing the distribution function with a density matrix. 

In this study, we investigate the scenario in which a quantum system has no equilibrium state to approach. Instead of relaxing to an equilibrium state, the system is in contact with two different baths such that it only approaches the nonequilibrium steady state (NESS), sustained by constant energy injections from the baths. In this case, we show that the parameter window for observing MPEs can be dramatically widened. Furthermore, we study the evolution of concurrence and quantum mutual information and investigate the MPEs in its dynamics. Lastly, we discuss the role quantum coherence plays in the system dynamics by contrasting simulations with and without the quantum coherence.

\begin{figure}[t]
\centering
\includegraphics[width=.9\columnwidth]{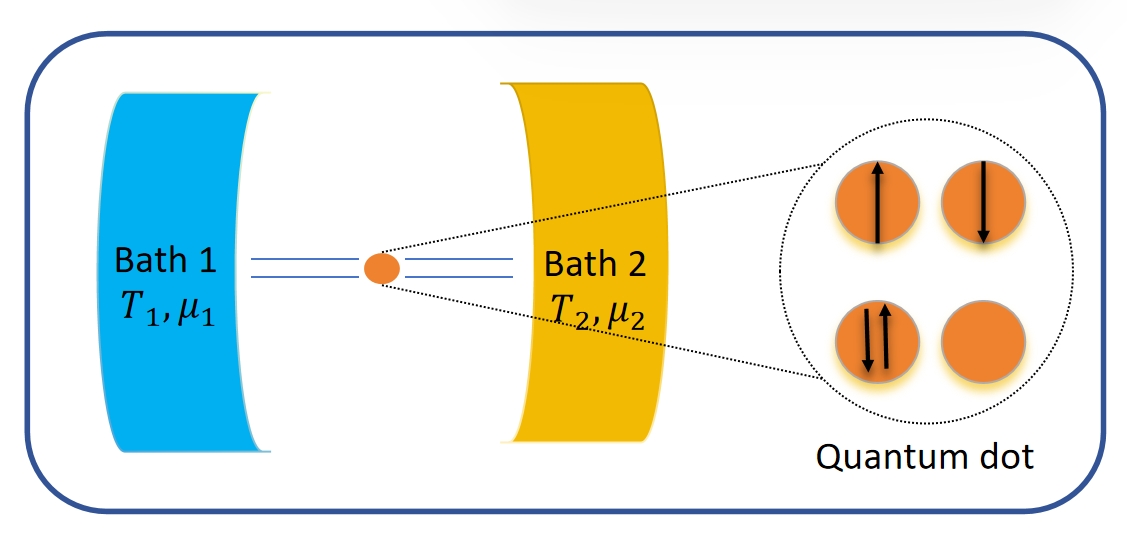}
\caption{Schematic of the quantum dot model between two baths. The quantum dot system initially prepared at certain steady state is put into the contact with the two baths at time $t=0$. The zoomed-in view of the system shows four different configurations of the quantum dot.}
\label{fig:0}
\end{figure}

\section*{Hamiltonians and Equations of Motion}
\subsection*{The quantum dot system}
To demonstrate the nonequilibrium boost to the MPEs in quantum dot system, we consider the quantum dot system coupled with two baths described by the Hamiltonian $\calh=\calh_S+\calh_R+\calh_{int}$ \cite{yang2011generic,eastham2013lindblad,qdot2023,harbola2006quantum,chatterjee2023multiple,yoshii2013analytical}:
\begin{eqnarray}
&&\calh_S=\sum_{\sigma}\epsilon_0 d^\dagger_\sigma d_\sigma+U n_\uparrow n_\downarrow\,, \cr
&&\calh_R=\sum_{k,\sigma}\omega_k \ a_{k\sigma}^\dagger a_{k\sigma}+\sum_{q,\sigma}\omega_q \ b_{q\sigma}^\dagger b_{q\sigma}\,,\cr
&&\calh_{int}= \sum_{k,\sigma} \lambda_k \left( d^\dagger_{\sigma}a_{k,\sigma}+ d^\dagger_{\sigma}b_{k,\sigma}\right)\,+\, \mathrm{h.c.}\,,
\label{eq:hdot} 
\end{eqnarray}
where $\calh_S,\, \calh_R,\, \calh_{int}$ represent the Hamiltonians for the system, the two reservoirs, and interaction between the system and the baths, respectively. $\epsilon_0$ is the electron energy in the quantum dot, $U$ is the repulsion energy of electrons, $n_{\sigma}=d_\sigma^\dagger d_\sigma$ is the number operator, and $\lambda_k$ is the reservoir-system coupling. Under the wide-band approximation (WBA), the density matrix of the system $\rho$ is of diagonal forms. The schematic illustration of the model is shown in Fig.~\ref{fig:0}.

We write the density matrix as a supervector $\rho_\alpha$ in the Liouville space representation where $\alpha=1,2,3,4$ represent states with double occupation, spin-up occupation, spin-down occupation, and the empty state. The system dynamics is described by the master equation 
\beq \frac{d}{dt}\rho_{i}(t) = \sum_{j} M_{ij}\rho_{j}(t)\,,  \label{Eq:meq}\eeq
where the transition matrix $M$ is given by \cite{yoshii2013analytical}
\begin{equation*}
M=\left(\begin{array}{cccc}
               -2(2-f^{(1)}) & f^{(1)} & f^{(1)} & 0 \\
               2-f^{(1)} & -2+f^{(0)}-f^{(1)} & 0 & f^{(0)} \\
               2-f^{(1)} & 0 & -2+f^{(0)}-f^{(1)} & f^{(0)} \\
               0 & 2-f^{(0)} &2-f^{(0)} & -2f^{(0)} \\
              \end{array}
\right).\end{equation*}   
Here, we have set the decay rate to $\Gamma=1$. $f^{(j)}$ with $j=0,1$ is defined as the sum of fermionic occupation numbers of the two baths $f^{(j)}=\dfrac{1}{1+e^{(\epsilon_0+jU-\mu_L)/T}}+\dfrac{1}{1+e^{(\epsilon_0+jU-\mu_R)/T}}$, where $\mu_{L(R)}$ represents the chemical potential of the left (right) bath. The time evolution of the density matrix element $\rho_\alpha(t)$ is given by
\begin{eqnarray}
\rho_\alpha(t)\,&=&\, \sum_{n=1}^{4} e^{\lambda_n t} {R}_{\alpha n}a_{n},
\label{eq:rho_alpha}
\end{eqnarray}
where $a_{n}=\,\sum_{m=1}^{4}{L}_{nm}\rho_{m}(0)$, $\lambda_n$ represents the eigenvalue of the transition matrix ${M}$, and $L_{ij}(R_{ij})$ represents the matrix of left (right) eigenvectors of the transition matrix given in the Supplementary Information.

\subsection*{The bipartite fermionic model}
An N-level quantum system truncated to two levels can be easily related to an anti-commuting fermionic system through the Jordan-Wigner transformation \cite{batista2001generalized}. To investigate the dynamics of quantum correlations, we consider a simple bipartite model with two sites. The system can be viewed as the two-site fermionic system with each site coupled to its own bath. Each site is either occupied by a fermion or vacant and the fermions can tunnel between the two sites. The Hamiltonians of the system  $\calh_S$, the two reservoirs $\calh_R$, and their mutual interactions $\mathcal{H}_{int}$ are given as follows:
\beq \begin{split}
&\calh_S= \sum_{i=1}^2 \omega_i \eta_i^\dagger \eta_i+\Delta(\eta_1^\dagger \eta_2+ \eta_2^\dagger \eta_1)\,, \\
&\calh_R=\sum_{k,p} \Omega_k \ (a_{kp}^\dagger a_{kp})+\sum_{q,s} \Omega_q \ (b_{qs}^\dagger b_{qs})\,,\\
& \mathcal{H}_{int}=\sum_{k,p}\lambda_k \ (\eta_1^\dagger a_{kp}+ \eta_1 a^\dagger_{kp})+\sum_{q,s} \lambda_q \ (\eta_2^\dagger b_{qs}+ \eta_2 b^\dagger_{qs})\,,
 \end{split}\eeq
where $\Delta$ is the tunneling rate between the two sites, $\eta_{i}^\dg$ is the fermionic creation operator on the $i$-th site, and $a_{kp}^{\dg} (b_{kp}^\dg)$ is the fermionic creation operator for a particle of momentum $k$, polarization $p$ in the reservoir. 
The quantum master equation for the density matrix of the two fermionic sites reads,
\beq \dot \rho(t)=\mathcal{L}[\rho(t)]=i[\rho(t),\mathcal{H}_S]-\sum_{i=1}^2N_i[\rho(t)]\,, \eeq
where $N_i[\rho]$ is the dissipation term into the $i$-th bath. The explicit form of the dissipator is given as follows:
\begin{align} 
N_i[\rho]&=\Gamma_1 \cdot \frac{1}{2}[1+(-1)^i \cos \theta] \ \left[(1- n_1^{T_i})(\zeta_1^\dagger \zeta_1 \tilde \rho - \zeta_1 \rho \zeta_1^\dagger) \right. \nonumber\\ 
& \hspace{3cm} \left. +n_1^{T_i} (\zeta_1 \zeta_1^\dagger \rho-\zeta_1^\dagger \rho \zeta_1)+h.c.  \right] \nonumber\\
&\quad +\Gamma_2 \cdot \frac{1}{2}[1+(-1)^{i-1} \cos \theta] \ \left[(1- n_2^{T_i})(\zeta_2^\dagger \zeta_2 \tilde \rho - \zeta_2 \rho \zeta_2^\dagger)\right.\\
&\hspace{3cm} \left.  +n_2^{T_i} (\zeta_2 \zeta_2^\dagger \rho-\zeta_2^\dagger \rho \zeta_2)+h.c.\right]\,,
\end{align}
where  \beq \vec{\zeta}=\begin{pmatrix} \cos(\theta/2) & \sin(\theta/2) \\ -\sin(\theta/2) & \cos(\theta/2) \end{pmatrix} \vec{\eta} \,,\eeq
cos$\theta=\dfrac{\omega_2-\omega_1}{\sqrt{(\omega_1-\omega_2)^2+4 \Delta^2}}$, $\omega_a'$ is the energy eigenvalue of the system, cos$\theta=\dfrac{w_2-w_1}{\sqrt{(w_1-w_2)^2+4 \Delta^2}}$, and $n_k^{T_i}$ is the number density of the $i^{\rm{th}}$ reservoir with temperature $T_i$ and energy $\omega^\prime_k$. 

For the symmetric case $\omega_1=\omega_2=1$, the eigenvalues of the transition matrix in the Lindblad equation only depend on the system decay rate $\Gamma$. To be specific,
\begin{gather}
    \lambda_1=-4\Gamma,\quad \lambda_2=\lambda_3=-2\Gamma,\quad \lambda_4=0\,,
\end{gather}
where we have assumed the equal decay rate of the two sites $\Gamma_1=\Gamma_2=\Gamma$. Then, the evolution of the system follows the trajectory which can be generically characterized by the following vectorized density operator equation 
\begin{gather}
    \Vec{\rho}(t)=\sum_i e^{\lambda_i t} \alpha_i(0) \Vec{v}_i\,,
\end{gather}
where $\Vec{v}_i$ is the $i$-th eigenvector of the transition matrix and $\alpha_i(0)$ is the coefficient of the $i$-th eigenvector defined by $\alpha_i(0)=\langle \Vec{\delta}_i,\Vec{\rho}(0)\rangle$. The matrix of vectors $\{\Vec{\delta}_i\}$ is the inverse of the matrix of eigenvectors $\{\Vec{v}_i\}$.

When taking coherence into account, we use the Redfield equation without ignoring the coherence terms. In this case, the quantum master equation has an additional coherence contribution, viz.,
\beq \dot \rho_S(t)=i[\rho(t),\mathcal{H}_S]-\sum_{i=1}^2N_i[\rho(t)]-\sum_{i=1}^2 S_i[\rho]\,, \eeq
where $S_i[\rho]$ describes the coherence exchange with the $i$-th bath and is given as follows:
\begin{align}
S_i[\rho]&=(-1)^{i-1} \frac{1}{2} \Gamma_1 \sin \theta \ \left[(1 - n_1^{T_i})(\zeta_2^\dagger \zeta_1 \rho  -  \zeta_1 \rho \zeta_2^\dagger) \right. \nonumber \\
 & \hspace{3cm} \left. + n_1^{T_i} (\zeta_2 \zeta_1^\dagger \rho- \zeta_1^\dagger \rho \zeta_2 )+h.c.\right] \nonumber\\
& \quad + (-1)^{i-1} \frac{1}{2} \Gamma_2 \sin\theta \ \left[(1 - n_2^{T_i})(\zeta_1^\dagger \zeta_2 \rho  - \zeta_1 \rho \zeta_2^\dagger ) \right.\nonumber\\
& \hspace{3cm} \left. +n_2^{T_i} (\zeta_1 \zeta_2^\dagger \rho - \zeta_1^\dagger \rho \zeta_2 )+h.c. \right] \,.
\end{align}

Usually, to have the ``strong Mpemba effect" \cite{klich2019mpemba,carollo2021exponentially}, the coefficient of the slowest decaying mode is required to vanish for the particularly chosen initial conditions. In this model, both $\alpha_2(0)$ and $\alpha_3(0)$ need to vanish due to the degeneracy of the eigenvalue. This gives the condition $\langle \Vec\delta_2,\Vec\rho(0)\rangle=\langle \Vec\delta_3,\Vec\rho(0)\rangle=0$. Notably, the strong MPE refers to the exponentially faster equilibration towards the equilibrium state, but does not guarantee the emergence of the anomalous crossing. For the nonequilibrium systems with multiple baths, a similar ``strong Mpemba effect" can be defined for systems that show exponentially faster decay to the NESS with vanishing coefficients for the slowest decaying mode. Generally, it is not necessary for a system to satisfy the strong Mpemba condition to manifest crossings of interested physical quantities during the relaxation. 

\subsection*{Quantum Correlations}
To demonstrate the dynamics of quantum correlations, we compute the concurrence and the quantum mutual information (QMI) between the two subsystems. The concurrence is a entanglement monotone derived from the entanglement formation \cite{hill1997entanglement,horodecki2009quantum}. For the system we consider, it can be simplified to
\begin{gather}
    \mathcal{E}(\rho)=2\max (0,\,|\rho_{23}^l|-\sqrt{\rho_{11}^l \rho_{44}^l})\,,
\end{gather}
where $\rho_{ij}^l$ are the entries of the density matrix in local basis. QMI is a direct generalization of classical mutual information and quantifies the maximum amount of information that can be securely transferred between two parties \cite{schumacher2006quantum,modi2012classical}. For a bipartite system $AB$ and the subsystem $A$($B$) with the reduced density operator $\rho^{A(B)}=Tr_{B(A)}(\rho^{AB})$, the QMI is defined as 
\begin{eqnarray}
\mathcal{I} (\rho^{AB}) = S (\rho^A) + S (\rho^B) - S(\rho^{AB})\ , \label{QMI}
\end{eqnarray}
where $S(\rho) = - \mathrm{tr} \, ( \rho \, \log_2 \rho )$ is the von Neumann entropy. 

\section*{Results}

\subsection*{Nonequilibrium boost to the MPEs in quantum dot system}

\begin{figure}[t]
\centering
\includegraphics[width=.49\columnwidth]{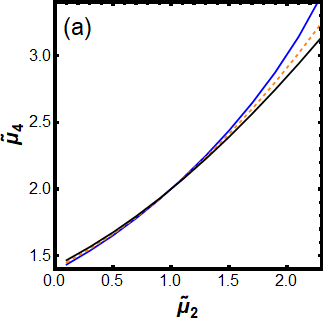}
\includegraphics[width=.47\columnwidth]{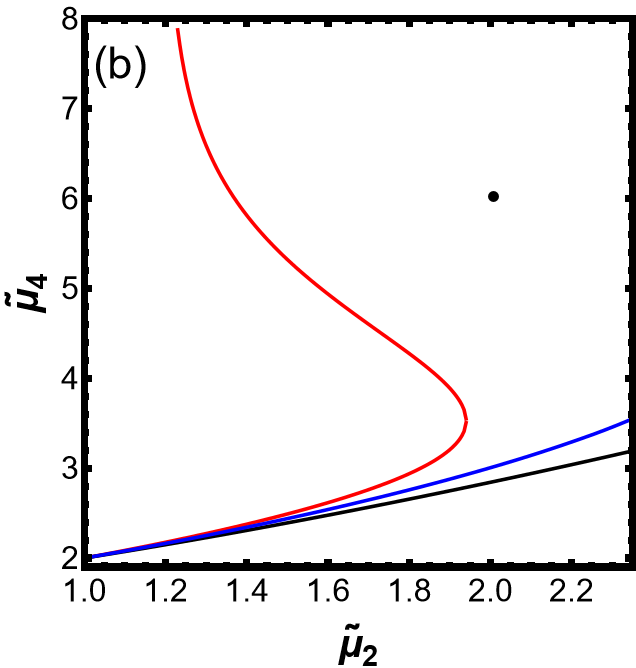}
\includegraphics[width=.98\columnwidth]{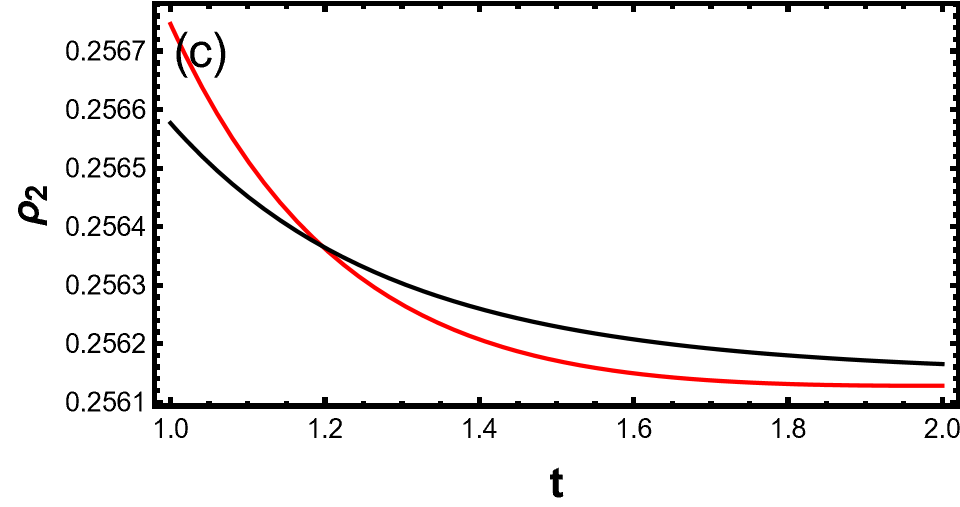}
\caption{(a) The viable range of $\tilde\mu_{2}$'s and $\tilde\mu_{4}$'s to induce QMPE in the equilibrium baths. The parameter regimes where QMPE emerges are between the black curve ($S_2=0$) and the blue curve ($S_2=-1$) intersecting at $(1,2)$. The dotted orange curve ($S_2=-0.2$) is inside the Mpemba regime. Here, $\mu_1=\mu_2=3$ in the unit of $k_B T$. (b) The boundaries of parameter regimes that show QMPE in the nonequilibrium case. In contrast to the narrow the Mpemba regime located between the blue and the black curves, the QMPE in nonequilibrium has a much further extended regime occupying the entire phase space between the black and the red curves. (c) Example of the evolution of the density matrix element $\rho_{2}$ at the $(\tilde\mu_2,\tilde\mu_4)=(2,6)$ indicated by the black dot in Fig.~\ref{fig:1} (b). Here, $\bar\mu=3,\,\Delta\mu=4,\,\tilde\mu_1=2,\,\tilde\mu_3=1$. For all the three graphs, the temperatures are set to be $k_B T_{1,2}=k_B \tilde T_{1,2,3,4}=k_B T$, and other parameters are set to be $\tilde\mu_1=2,\,\tilde\mu_3=1,\, \epsilon_0=2,\, U=1.25$ in the unit of $k_B T$.}
\label{fig:1}
\end{figure}

The investigated MPEs are mostly on the systems that are relaxing to their final equilibrium states with surrounding environments. Naively, it seems that the asymptotic final state does not influence the emergence of the MPE since the Mpemba crossing occurs during the process of relaxation before equilibration. This thinking can be simplistic. The final state is determined by both the system parameters and the environment. Given that the environment is a crucial driving force of the system dynamics, its condition can dramatically influence the trajectory of the system evolution as well as the emergence of the MPEs.

One indicator for the emergence of the MPE is the crossing of the density matrix elements. For the quantum dot system with the two initial states $\rho^\mathrm{I}(0)$ and $\rho^\mathrm{II}(0)$ prepared to be the steady states with initial preparing baths, the criteria for the emergence of Mpemba effect in the $n$-th elements of the density matrices was shown to take a simple analytic form $-1< S_n=\frac{R_{n,3}\Delta a_3}{R_{n,4}\Delta a_4}<0$ \cite{qdot2023}, where $R_{ij}$ represents the matrix of right eigenvectors of the transition matrix, and $\Delta a_i$ is the difference between the transformed density matrix elements of the two initial states defined by $a_{n}=\,\sum_{m}L_{n,m}\Delta\rho_{m}(0)$\footnote{The slowest decaying mode vanishes identically for all initial states in equilibrium with the environment, suggesting an exponential speedup in its decay compared to most randomly selected initial states. 
Notably, this speedup not only holds for the initial states in equilibrium with the baths, but also for the NESS sitting between two preparing baths of different temperatures or chemical potentials.}. 
Here the left eigenmatrix of the transition matrix $L_{n,m}$ in \eqref{Eq:meq} is defined by $LM=diag(\lambda_i)M$ where $\lambda_i$'s are the eigenvalues of the transition matrix $M$. The emergence of MPE in the full and empty states depends mostly on the initial conditions, while the dynamics of the population on the singly-occupied states respond sensitively to the environmental changes. Therefore, we focus on the population of the singly-occupied spin-up state $\rho_2$ as an observable to demonstrate the MPEs. In the Letter, quantities associated with the preparing baths used to prepare the initial states of the system are indicated by the tildes over the variables. For example, $\tilde\mu_{1(2)}$ and $\tilde\mu_{3(4)}$ represent the chemical potentials of the two pairs of preparing baths for initial states $\rho^\mathrm{I}(0)$ and $\rho^\mathrm{II}(0)$, respectively. In case of two identical baths, the parameter regimes where the population of the spin-up state $\rho_2$ has the Mpemba crossing are shown in Fig.~\ref{fig:1} (a).The viable choices of $(\tilde\mu_{2},\tilde\mu_{4})$'s one can use to induce the QMPE is within the phase space bounded by the black and the blue curves intersecting at $(\tilde\mu_{2},\tilde\mu_{4})=(1,2)$.

\begin{figure}[t]
\centering
\includegraphics[width=.49\columnwidth]{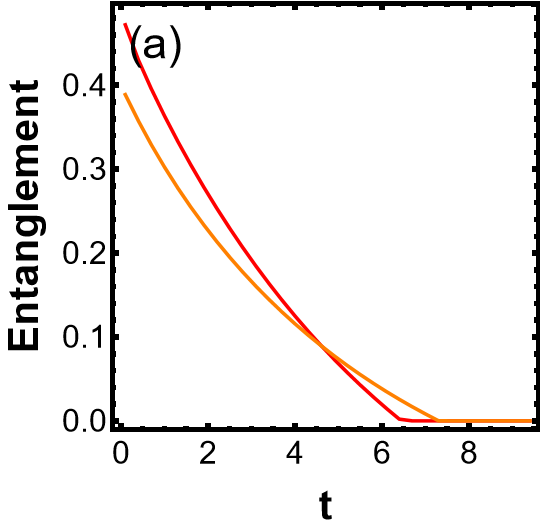}
\includegraphics[width=.47\columnwidth]{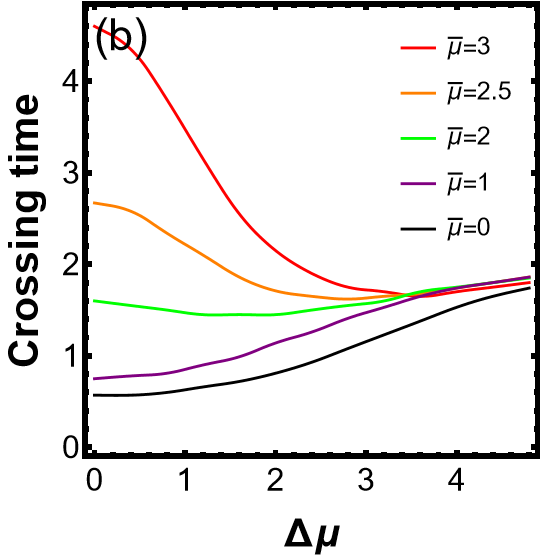}
\caption{(a) Evolution of entanglement between the two sites as a function of time. The red curve is for the initial state $|\rho^\mathrm{I}(0)\rangle=\{0,0.2,0.7,0.1\}$ in the energy eigenbasis of $\calh_S$ and the orange curve represents the initial state $|\rho^\mathrm{II}(0)\rangle=\{0.1,0.7,0.1,0.1\}$. Here, $\mu_1=\mu_2=3$. (b) The time of Mpemba crossing as a function of chemical potential bias $\Delta \mu$. Here, $\mu_1=\bar\mu+\Delta\mu,\,\mu_2=\bar\mu-\Delta\mu$. For both, the parameters used are: $T_1=T_2=T,\,\Gamma=0.05,\,\omega_1=\omega_2=1,\,\Delta=0.2$. The values of the parameters are in the unit of $k_B T$.}
\label{fig:4}
\end{figure}

To investigate the MPEs in nonequilibrium environments, we vary the chemical potential bias $\Delta\mu$ between the two baths \footnote{The difference between the two initial states $\Delta a_3$ and $\Delta a_4$ are set to be constant. This procedure is possible for the density matrix elements $\rho_{1}$ and $\rho_2$ since their values are solely determined by the initial conditions.}. When moved from the equilibrium to the nonequilibrium regimes, the boundary of the Mpemba regime represented by the black curve  ($S_2=0$) in Fig.~\ref{fig:1} (a) remains unchanged, while the other boundary ($S_2=-1$) shifts dramatically. This is due to the fact that the matrix of the left eigenvectors $L$ of the transition matrix is not a full-rank matrix and satisfies the condition 
\begin{gather}
    L\ \Vec{v}=0, \quad \mathrm{where} \quad \Vec{v}=\{-1,1,1,-1\}^T \,.
\end{gather}
The equation $S_n=0$ is automatically satisfied when the vector of initial density matrix elements satisfies the condition $\Delta \Vec{\rho}_i \propto \{-1,1,1,-1\}^T$, which does not depend on the environmental parameters. Notably, the boundary shift diverges as $\Delta\mu$ goes beyond certain threshold $\Delta\mu^*$ \cite{sm}. This suggests that when strong nonequilibrium conditions are introduced, i.e., $\Delta\mu>\Delta\mu^*$, an arbitrary large chemical potential in the preparing bath $\tilde\mu_4$ will induce the nonequilibrium QMPE. Physically, this enhancement is due to the biased decaying rates caused by the two distinct reservoirs. As shown in Fig.~\ref{fig:1} (b), the equilibrium QMPE emerges within the regime bounded by the blue and the black curves, and the nonequilibrium condition of the two baths expands the Mpemba regime into one bounded by the red and the black curves. Interestingly, the red curve, which one of the boundaries of the Mpemba regime, ceases to extend beyond a finite $\tilde\mu_2$ and leaves the entire regimes on top of the black curve succumbed to the QMPE. This shows a massive extention in the phase space of the Mpemba regime and opens up a much broader parameter window for experimental investigations of the QMPE. An example of the QMPE at $(\tilde\mu_2,\tilde\mu_4)=(2,6)$ indicated by the black dot in  Fig.~\ref{fig:1} (b) is demonstrated in Fig.~\ref{fig:1} (c).

\subsection*{Emergence of MPEs in quantum correlations}

Quantum correlations are often viewed as resources for faster quantum processing and play a central role in the quantum thermodynamics and quantum computing \cite{chitambar2019quantum}. The dynamics of correlations have been shown to demonstrate peculiar features such as anomalous symmetry restorations and dynamic phase transitions  \cite{ares2023entanglement,vosk2014dynamical,de2021entanglement,skinner2019measurement,li2023high}. It is unclear if anomalous phenomena such as QMPEs can emerge dynamically.

\begin{figure}[t]
\centering
\includegraphics[width=0.49\columnwidth]{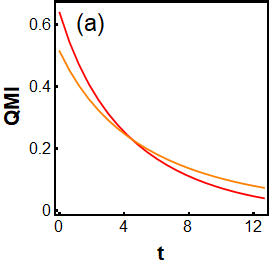}
\includegraphics[width=0.49\columnwidth]{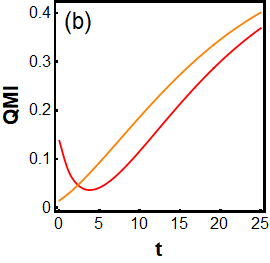}
\caption{MPE and inverse MPE in the evolutions of the QMI between the two sites. (a) The red curve represents the evolution of the initial state $|\rho^\mathrm{I}(0)\rangle=\{0.1,0.1, 0.7, 0.1\}$ in the energy eigenbasis of $\calh_S$ and the orange curve represents the initial state $|\rho^\mathrm{II}(0)\rangle=\{0.1, 0.65, 0.1, 0.15\}$. Parameters used are: $T_1=T_2=T,\,\mu_1=\mu_2=3$. (b)  The red curve represents $|\rho(0)\rangle=\{0.4,0.1, 0.2, 0.3\}$ and the orange curve is for $|\rho(0)\rangle=\{0.3, 0.3, 0.2, 0.2\}$. Parameters used are: $T_1=T_2=0.1,\,\mu_1=\mu_2=1.2$. For both, $\Gamma=0.05,\,\omega_1=\omega_2=1,\,\Delta=0.2$. The values of the parameters are in the unit of $k_B T$.}
\label{fig:5}
\end{figure}

The dynamics of the entanglement and the QMI for our system are shown in Fig.~\ref{fig:4} and \ref{fig:5}. In Fig.~\ref{fig:4}, we demonstrate the emergence of the QMPE in the entanglement evolution from two different initial states. The initially more entangled state, represented by the red curve in Fig.~\ref{fig:4} (a), has a faster rate of disentanglement and untangles earlier than the state initially less entangled. This represents possible tradeoffs between the entangling time and the entangling strength for the quantum states. The vanishing of entanglement within a finite time is termed the ``sudden death" in contrast to the smooth asymptotic decay observed in the dynamics of QMI \cite{yu2009sudden}. The nonequilibrium conditions can significantly influence the crossing time of the entanglements and advance it when the average chemical potential $\bar\mu$ of the baths is large [see Fig.~\ref{fig:4} (b)]. It can be intuitively understood as follows. When the baths are set at large $\mu$'s, the quantum states are highly occupied. Increasing the bias in this case means lowering the occupation of one of the bath while keeping the occupation of the other bath roughly unchanged. In this case, the state $\rho^\mathrm{I}(0)$ which has a higher excitation energy obtains a higher rate of decay, resulting in an earlier crossing time. On the other hand, for very low chemical potentials the fermion states are approximately vacant. Enlarging $\Delta\mu$ effectively raises the occupation number of one of the baths. This increase in the bath causes a slower decay and a delayed crossing time. Similar arguments explains why the crossing time approaches a constant for extremely large $\Delta\mu$'s regardless of $\bar\mu$'s. In Fig.~\ref{fig:5}, we show the MPE and the inverse MPE in the evolution of the QMI. These two figures jointly demonstrate the ubiquity of Mpemba crossings in the evolution of quantum systems.

\begin{figure}[t]
\centering
\includegraphics[width=0.49\columnwidth]{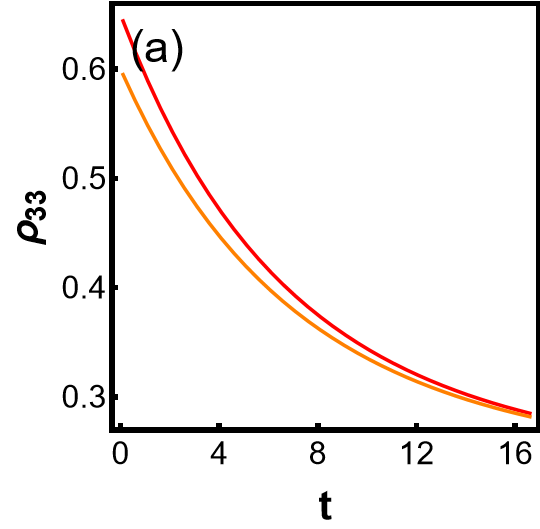}
\includegraphics[width=0.49\columnwidth]{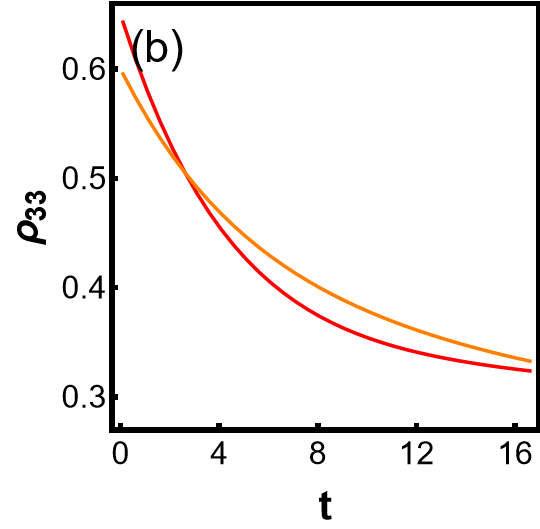}
\includegraphics[width=0.99\columnwidth]{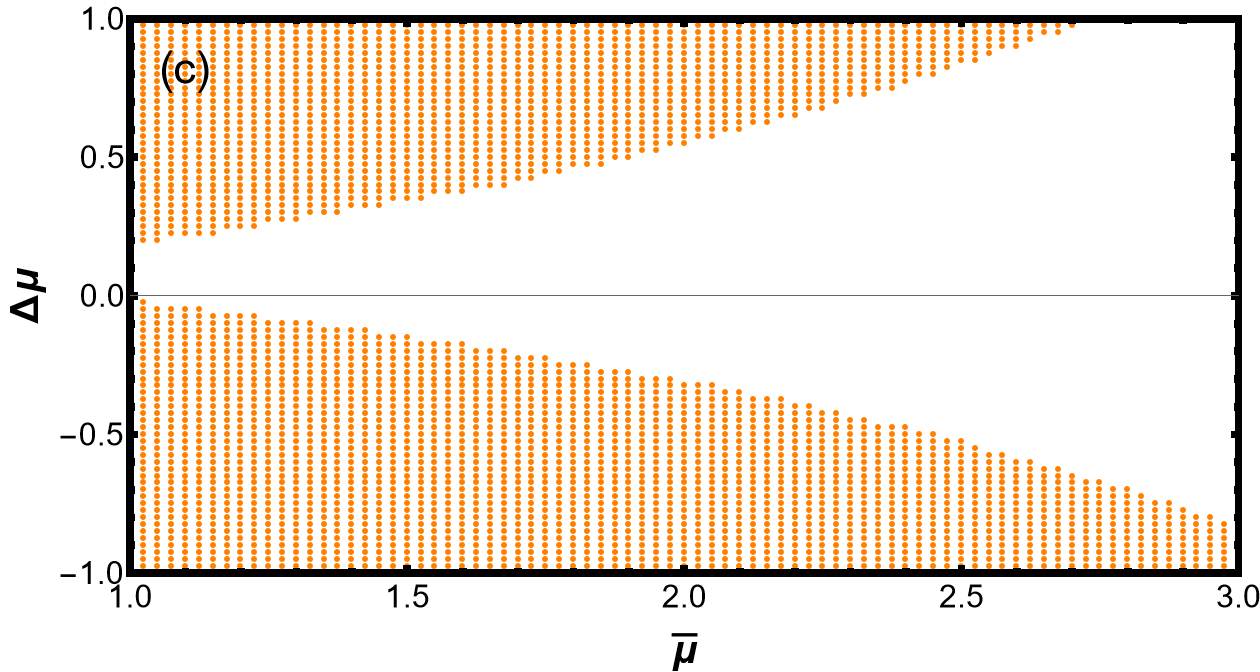}
\caption{Evolution of the density matrix element $\rho_{33}$ as a function of time (a) with and (b) without coherence terms. The off-diagonal terms in the initial condition are chosen at $\rho^\mathrm{I}_{23}(0)=\rho^\mathrm{I}_{32}(0)=0.2$ for the red curve and $\rho^\mathrm{II}_{23}(0)=\rho^\mathrm{II}_{32}(0)=-0.1$ for the orange curve. For both (a) and (b), the population terms are identical. The red curve represents the the initial state with the population terms $|\rho^\mathrm{I}(0)\rangle=\{0.1,0.25, 0.65, 0\}$ and the orange curve represents $|\rho^\mathrm{II}(0)\rangle=\{0.1, 0.2, 0.6, 0.1\}$. (c) The shaded region in orange represents the regime for the MPE of (b) with $\mu_{1,2}=\bar\mu\pm\Delta\mu$. Parameters used are: $T_1=T_2=1,\,\mu_1=0.1,\,\mu_2=3,\,\Gamma=0.05,\,\omega_1=\omega_2=1,\,\Delta=0.05$. }
\label{fig:6}
\end{figure}

\subsection*{Influence from the quantum coherence}


One of the essential features of a quantum system is the off-diagonal coherence, which distinguishes its density matrix from a classical one \cite{streltsov2017colloquium}. In the conventional Lindbladian treatment, the coherence terms represented by the off-diagonal elements in the density matrix decouple from the population terms in the dynamics. Under the WBA used in quantum dot systems, the coherence terms are erased completely. For a quantum system out of equilibrium and in contact with multiple baths, these usually ignored coherence can be much amplified especially when the tunneling rate is comparable to the decoherence rate ($\Delta \lesssim 2 \Gamma$) \cite{wang2019nonequilibrium,wang2022effect,zhang2021quantum}. We show that the dynamics with quantum coherence can lead to the emergence of QMPE while the population dynamics without it predict the otherwise. This distinctive feature caused by coherence can only take place when nonequilibrium conditions are introduced. 

In Fig.~\ref{fig:6} (a) and (b), we show that in the regime where no sign of the QMPE is witnessed when coherence is ignored, it emerges when coherence is incorporated even when the initial states are similarly prepared without coherence. The coupling between the coherence and the population terms enables the coherence to be dynamically generated and create influence back on the population terms of the system. Such dramatic effect only emerges when strong nonequilibrium conditions are set up to drive the system away from equilibrium distributions. For larger average potentials $\bar\mu$, a larger $\Delta\mu$ is required for the emergence the QMPE [see Fig.~\ref{fig:6} (c)]. This is due to the fact that larger biases are necessary to substantially alter the occupation number of the baths at high chemical potentials. Small biases fail to create enough differences from the equilibrium solutions, consequently, do not generate enough coherence to trigger the QMPE. This is a demonstration of the quantum coherence causing striking difference on the dynamics of the system comparable to a phase transition. Importantly, the magnitude of the quantum coherence is amplified in the weak tunneling regimes and its asymptotic value is approximately proportional to the bias between the baths occupation numbers \cite{wang2019nonequilibrium}. Therefore, this ``coherence-induced MPE" is an intrinsic nonequilibrium phenomenon that only emerges when the multiple nonequilibrium supporting baths of subsystems are introduced.

\section*{Discussion and Conclusion}
Though discovered half a century ago, observation of the Mpemba effect has been difficult due to stringent experimental conditions required. In this study, we found that in out-of-equilibrium systems without asymptotic equilibrium states, such effect in quantum realm is ubiquitous. The out-of-equilibrium environments have dramatic influence on the emergence of the Mpemba effect, significantly broadening the range of the Mpemba regime, expediting its emergence and inducing sufficient quantum coherence to trigger the effect. This may point to a promising direction for future experimental investigations.

In this study, we investigated the quantum Mpemba effect in the quantum dot and the two-site fermion systems coupled with two different baths. We showed that nonequilibrium conditions can dramatically expand the parameter space where the MPE emerges. This opens up a much wider window both conceptually and also practically for experimental investigations. However, the Mpemba effect defined in the quantum dot system was investigated through the population terms of the density matrix which represent certain spin configurations of the system. We did not find similar crossings in the relative entropies between the prepared systems and the final state. It is unclear if such effect can be extended to the distance measures such as relative entropy or fidelity for the system using the Markovian Lindblad equation. it is worth noticing that the similar effect was found when non-Markovian effects are taken into account \cite{strachan2024non}. Besides the population dynamics, we also invetigated the anomalous decays of MPEs and inverse MPEs emerge in the evolution of the entanglement and the QMI in the two-fermionic system coupled with two different baths. We explained that nonequilibrium conditions can significantly influence the times of the Mpemba crossings. Besides, we studied the possible influence on the dynamics due to the quantum coherence sustained by nonequilibrium which is absent in the conventional Lindbladian dynamics. We conduct simulations both with and without quantum coherence and showed show that the quantum coherence supported by the nonequilibrium conditions of the two baths has qualitative influence the population dynamics and can induce the emergence of the QMPE which the population dynamics alone fail to predict.

\begin{widetext}

\section*{Supplemental Material}
In the Supplemental Material, we provide certain calculation details of the quantum dot and the two fermion models in the main text as well as extra figures on the MPEs in the quantum dot system.

\subsection{MPE in the quantum dot model}

The Lindblad equation of the quantum dot model has been studied in many previous papers \cite{eastham2013lindblad,qdot2023,harbola2006quantum}. The results we used in this study are summarized in this section which can be found in Refs.~\cite{qdot2023,chatterjee2023multiple,yoshii2013analytical}. The quantum dot system coupled with two baths is described by the total Hamiltonian $\calh=\calh_S+\calh_R+\calh_{int}$ given as follows \cite{yang2011generic}:
\begin{eqnarray}
&&\calh_S=\sum_{\sigma}\epsilon_0 d^\dagger_\sigma d_\sigma+U n_\uparrow n_\downarrow\,, \cr
&&\calh_R=\sum_{k,\sigma}\omega_k \ a_{k\sigma}^\dagger a_{k\sigma}+\sum_{q,\sigma}\omega_q \ b_{q\sigma}^\dagger b_{q\sigma}\,,\cr
&&\calh_{int}= \sum_{k,\sigma} \lambda_k \left( d^\dagger_{\sigma}a_{k,\sigma}+ d^\dagger_{\sigma}b_{k,\sigma}\right)\,+\, \mathrm{h.c.}\,,
\label{eq:hdot} 
\end{eqnarray}
where $\calh_S,\, \calh_R,\, \calh_{int}$ represent the Hamiltonians for the system, the two reservoirs, and interaction between the system and the baths, respectively. $\epsilon_0$ is the electron energy in the quantum dot, $U$ is the repulsion energy of electrons, $n_{\sigma}=d_\sigma^\dagger d_\sigma$ is the number operator, and $\lambda_k$ is the reservoir-system coupling. The creation (annihilation) operators $d^\dagger_\sigma\,(d_\sigma)$ and $a^\dagger_\sigma\,(a_\sigma)$ follow the fermionic statistics $\{d_\alpha,d^\dagger_{\beta}\}=\delta_{\alpha\beta}$. The four spin configurations -- the doubly-occupied state, two singly-occupied states, and empty state -- are labeled by $\alpha=1,2,3,4$, respectively. The quantum master equation in the interaction picture of the system density matrix can be written as
\begin{align}
    \frac{d}{dt}\rho_s=-\Tr_{L,R} \int_0^t ds [\calh_{int}(t),[\calh_{int}(s),\rho_s(t)\otimes \rho_L\otimes \rho_R]]\,,
\end{align}
where $L (R)$ represents the left (right) bath. Under the wide-band approximation, the quantum master equation of the vectorized density matrix elements is given by 
\begin{equation}
\frac{d}{d t}{\rho_i}=\sum_j{M_{ij}}{\rho_j},
\label{eq:qme} 
\end{equation}
where the transition matrix is \cite{yoshii2013analytical}
\begin{equation}
{M_{ij}}=\left(\begin{array}{cccc}
               -2(2-f^{(1)}) & f^{(1)} & f^{(1)} & 0 \\
               (2-f^{(1)}) & -(2-f^{(0)})-f^{(1)} & 0 & f^{(0)} \\
               (2-f^{(1)}) & 0 & -(2-f^{(0)})-f^{(1)} & f^{(0)} \\
               0 & (2-f^{(0)}) &(2-f^{(0)}) & -2f^{(0)} \\
              \end{array}
\right).
\end{equation}
Here, we have set the decay rate to $\Gamma=1$. $f^{(j)}$ with $j=0,1$ is defined as the sum of fermionic occupation numbers of the two baths $f^{(j)}=\dfrac{1}{1+e^{(\epsilon_0+jU-\mu_L)/T}}+\dfrac{1}{1+e^{(\epsilon_0+jU-\mu_R)/T}}$, where $\mu_{L(R)}$ represents the chemical potential of the left (right) bath. The time evolution of the density matrix element $\rho_\alpha(t)$ is given by
\begin{eqnarray}
\rho_\alpha(t)\,&=&\, \sum_{n=1}^{4} e^{\lambda_n t} {R}_{\alpha n}a_{n},
\label{eq:rho_alpha}
\end{eqnarray}
where $a_{n}=\,\sum_{m=1}^{4}{L}_{nm}\rho_{m}(0)$, $\lambda_n$ represents the eigenvalue of the transition matrix ${M}$, and $L_{ij}(R_{ij})$ represents the matrix of left (right) eigenvectors of the transition matrix given as follows:
\begin{equation}
{R}=\left(\begin{array}{cccc}
\frac{f^{(0)}f^{(1)}}{4+2(f^{(0)}-f^{(1)})} & 0 & \frac{2f^{(0)}f^{(1)}}{-4+\left(f^{(0)}-f^{(1)}\right)^2} & \frac{f^{(0)}f^{(1)}}{4-2(f^{(0)}-f^{(1)})}\\
\frac{f^{(0)}(2-f^{(1)})}{4+2(f^{(0)}-f^{(1)})} & -\frac{1}{2}  & -\frac{f^{(0)}(2-f^{(0)}-f^{(1)})}{-4+\left(f^{(0)}-f^{(1)}\right)^2} & -\frac{f^{(0)}f^{(1)}}{4-2(f^{(0)}-f^{(1)})}\\
\frac{f^{(0)}(2-f^{(1)})}{4+2(f^{(0)}-f^{(1)})} & \frac{1}{2}   & -\frac{f^{(0)}(2-f^{(0)}-f^{(1)})}{-4+\left(f^{(0)}-f^{(1)}\right)^2} & -\frac{f^{(0)}f^{(1)}}{4-2(f^{(0)}-f^{(1)})}\\
\frac{(2-f^{(0)})(2-f^{(1)})}{4+2(f^{(0)}-f^{(1)})} & 0  & -\frac{2f^{(0)}(2-f^{(0)})}{-4+\left(f^{(0)}-f^{(1)}\right)^2} & \frac{f^{(0)}f^{(1)}}{4-2(f^{(0)}-f^{(1)})}\\
\end{array}
\right),
\end{equation}
and
\begin{equation}
{L}=\left(\begin{array}{cccc}
               1 & 1 & 1 & 1 \\
               0 & -1 & 1 & 0 \\
               -\frac{2-f^{(1)}}{f^{(0)}} & -\frac{2-f^{(0)}-f^{(1)}}{2f^{(0)}} & -\frac{2-f^{(0)}-f^{(1)}}{2f^{(0)}} & 1 \\
               \frac{(2-f^{(0)})(2-f^{(1)})}{f^{(0)}f^{(1)}} & -\frac{2-f^{(0)}}{f^{(0)}} & -\frac{2-f^{(0)}}{f^{(0)}} & 1 \\
              \end{array}
\right).
\end{equation}

\begin{figure}[t]
\centering
\includegraphics[width=.39\columnwidth]{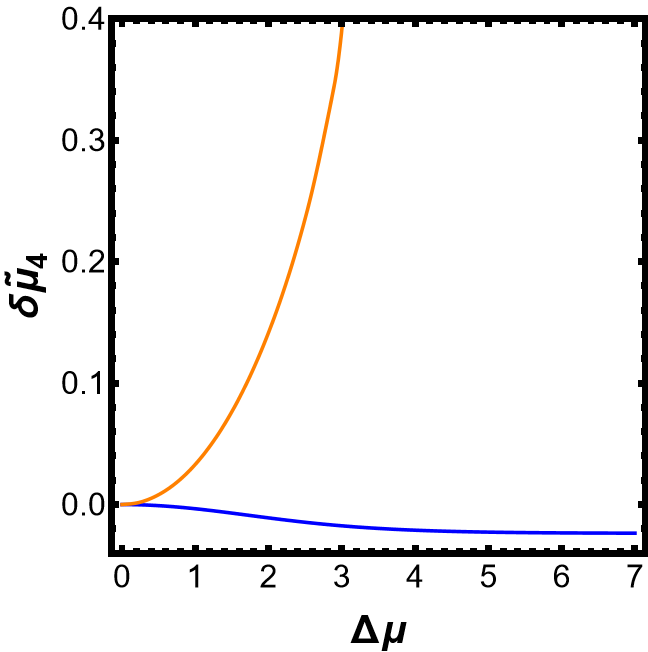}\\%
\caption{Shifts of the blue boundary ($S_2=-1$) in Fig.~1 (a) in the main text due to nonequilibrium baths. The chemical potential of the left preparing bath is fixed at $\tilde\mu_2=0$ (blue curve) and $\tilde\mu_2=2$ (orange curve). The two baths that the system is in contact with during relaxation are set to $\mu_1=3-\Delta \mu,\,\mu_2=3+\Delta\mu$. Parameters used are: $T_{1,2}=\tilde T_{1,2,3,4}=1,\,\tilde\mu_1=2,\,\tilde\mu_3=1,\, \epsilon_0=2,\, U=1.25$.}
\label{fig:1}
\end{figure}

For different $\tilde\mu_2$'s, the boundary shift of the Mpemba regime has opposite trends before and after the intercepting point $\tilde\mu_2=1$ as $\Delta\mu$ enlarges -- it shifts downwards for $\tilde\mu_2<1$ and shift upwards for $\tilde\mu_2>1$. An example demonstrating the shifts of the boundary is shown in Fig.~\ref{fig:1} where the orange curve is for $\tilde\mu_2=2$ and the blue curve is for $\tilde\mu_2=0$. The boundary shift diverges as $\Delta\mu$ goes beyond certain threshold. One can numerically determine that the threshold value of the chemical potential bias is $\Delta\mu^*\approx 3.2$ in the example. This suggests that when strong nonequilibrium conditions are introduced, i.e., $\Delta\mu>\Delta\mu^*$, an arbitrary large $\tilde\mu_4$ will induce the nonequilibrium QMPE.


In Fig.~\ref{fig:ap2}, we show that the vast extension of parameter space of the Mpemba regime also exists for the other singly-occupied state $\rho_3$. For the doubly-occupied and vacant states, the nonequilibrium condition has no influence on the Mpemba regime.

\begin{figure}[h]
\centering
\includegraphics[width=19pc]{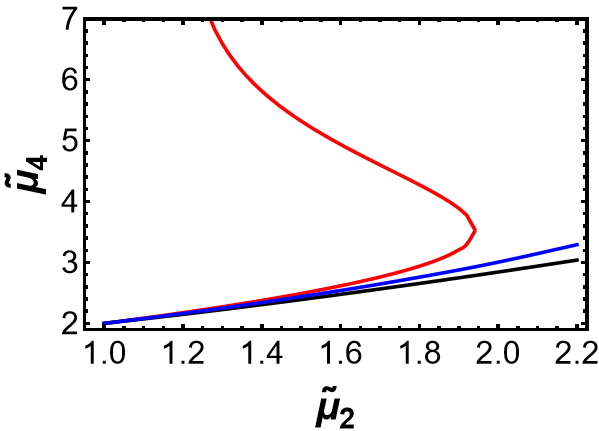}
\caption{The boundaries of parameter regime that show QMPE for the third density matrix element $\rho_3$ ($-1<S_3<0$). The Mpemba regime in the equilibrium environment expands between the blue and the black curves, and the QMPE in nonequilibrium extends to the parameter space between the black and the red curves. Parameters used are $\bar\mu=3,\,\Delta\mu=4,\,\tilde\mu_1=2,\,\tilde\mu_3=1$. For all, $T_1=T_2=\tilde T_{1}=\tilde T_{2}=\tilde T_{3}=\tilde T_{4}=1,\,\tilde\mu_1=2,\,\tilde\mu_3=1,\, \epsilon_0=2,\, U=1.25$.}
\label{fig:ap2}
\end{figure}

\subsection{MPE in the two fermion model}

The quantum master equation including the coherence terms for the density matrix of the two fermionic sites is,
\beq \dot \rho_S(t)=i[\rho_S,H_S]-D_0[\rho]-D_s[\rho], \label{mst} \eeq
where the dissipators describing the particle exchange and the coherence exchange with the $i$-th bath are
\beq D_0[\rho]=\sum_{i=1}^2N_i[\rho], \ \ D_s[\rho]=\sum_{i=1}^2 S_i[\rho]\, . \eeq
$N_i[\rho]$ and $S_i[\rho]$ are defined as follows,
\begin{equation} 
\begin{split}
    N_i[\rho]&=\Gamma_1 \cdot \frac{1}{2}[1+(-1)^i \cos \theta] \ \left[(1- n_1^{T_i})(\zeta_1^\dagger \zeta_1 \tilde \rho - \zeta_1 \rhot \zeta_1^\dagger) +n_1^{T_i} (\zeta_1 \zeta_1^\dagger \rhot-\zeta_1^\dagger \rhot \zeta_1)+h.c. \right] \\
    &+\Gamma_2 \cdot \frac{1}{2}[1+(-1)^{i-1} \cos \theta] \ \left[(1- n_2^{T_i})(\zeta_2^\dagger \zeta_2 \tilde \rho - \zeta_2 \rhot \zeta_2^\dagger) +n_2^{T_i} (\zeta_2 \zeta_2^\dagger \rhot-\zeta_2^\dagger \rhot \zeta_2)+h.c.\right],
\end{split}
\end{equation}
and
\begin{align}
S_i[\rho]&=(-1)^{i-1} \frac{1}{2} \Gamma_1 \sin \theta \, \left[(1 - n_1^{T_i})(\zeta_2^\dagger \zeta_1 \rhot  -  \zeta_1 \rhot \zeta_2^\dagger) + n_1^{T_i} (\zeta_2 \zeta_1^\dagger \rhot- \zeta_1^\dagger \rhot \zeta_2 )\right] \nonumber\\
& + (-1)^{i-1} \frac{1}{2} \Gamma_2 \sin\theta \, \left[(1 - n_2^{T_i})(\zeta_1^\dagger \zeta_2 \rhot  - \zeta_1 \rhot \zeta_2^\dagger ) +n_2^{T_i} (\zeta_1 \zeta_2^\dagger \rhot - \zeta_1^\dagger \rhot \zeta_2) \right]+h.c. \,,
\end{align}
where  \beq \vec{\zeta}=\begin{pmatrix} \cos(\theta/2) & \sin(\theta/2) \\ -\sin(\theta/2) & \cos(\theta/2) \end{pmatrix} \vec{\eta} \,,\eeq
cos$\theta=\dfrac{\omega_2-\omega_1}{\sqrt{(\omega_1-\omega_2)^2+4 \Delta^2}}$, $\omega_a'$ is the energy eigenvalue of the system, cos$\theta=\dfrac{w_2-w_1}{\sqrt{(w_1-w_2)^2+4 \Delta^2}}$, and $n_k^{T_i}$ is the number density of the $i^{\rm{th}}$ reservoir with temperature $T_i$ and energy $\omega^\prime_k$.

In the energy eigenbasis, the master equation for the density matrix was calculated to be
\begin{equation}
    \frac{d}{dt} \rho_{ij}=\sum_{kl} M_{ij}^{kl}\rho_{kl}\,,
\end{equation}
where the nonzero matrix elements $M_{ij}^{lk}$ for fermionic reservoirs without consideration of the quantum coherence terms are given as follows,
\begin{align}
M_{11}^{11}&=-2(\Gamma_1 (\sin^2(\theta/2)\, n_1^{T_1}+\cos^2(\theta/2)\, n_1^{T_2})+\Gamma_2 (\cos^2(\theta/2)\, n_2^{T_1}+\sin^2(\theta/2)\, n_2^{T_2})),\\
M_{11}^{22}&=-2\Gamma_1 (\sin^2(\theta/2)\, n_1^{T_1}+\cos^2(\theta/2)\, n_1^{T_2})+2\Gamma_1,\\
M_{11}^{33}&=-2\Gamma_2 (\cos^2(\theta/2)\, n_2^{T_1}+\sin^2(\theta/2)\, n_2^{T_2})+2\Gamma_2,\\
M_{22}^{11}&=-M_{11}^{22}+2\Gamma_1, \\
M_{22}^{22}&=-M_{11}^{22}+M_{11}^{33}-2\Gamma_2,\\
M_{22}^{44}&=M_{11}^{33}, \\
M_{33}^{11}&=2\Gamma_2 (\cos^2(\theta/2)\, n_2^{T_1}+\sin^2(\theta/2)\, n_2^{T_2}),\\
M_{33}^{33}&=M_{11}^{22}-2\Gamma_1-M_{11}^{33},\\
M_{33}^{44}&=M_{11}^{22},\\
M_{44}^{22}&=M_{33}^{11},\\
M_{44}^{33}&=M_{22}^{11},\\
M_{44}^{44}&=-M_{11}^{22}-M_{11}^{33},
\end{align}
where cos$\theta=\dfrac{\omega_2-\omega_1}{\sqrt{(\omega_1-\omega_2)^2+4 \Delta^2}}$. 

The time evolution of the vectorized density operator is given by 
\begin{gather}
    \Vec{\rho}(t)=\sum_i e^{\lambda_i t} \alpha_i(0) \Vec{v}_i\,,
\end{gather}
where $\Vec{v}_i$ is the $i$-th eigenvector of the transition matrix and $\alpha_i(0)$ is the coefficient of the $i$-th eigenvector defined by $\alpha_i(0)=\langle \Vec{\delta}_i,\Vec{\rho}(0)\rangle$. The matrix of vectors $\{\Vec{\delta}_i\}$ is the inverse of the matrix of eigenvectors $\{\Vec{v}_i\}$. The matrix of the row vectors of $\Vec{\delta}_i$'s is given by
\begin{eqnarray}
    \left(
\begin{array}{cccc}
 \frac{1}{4} n_{1 p} n_{2 p} & \frac{1}{4} \left(n_{1 p}-2\right) n_{2 p} & \frac{1}{4} n_{1 p} \left(n_{2 p}-2\right) & \frac{1}{4} \left(n_{1 p}-2\right) \left(n_{2 p}-2\right) \\
 -\frac{1}{2} \left(n_{1 p} n_{2 p}\right) & -\frac{1}{2} \left(n_{1 p}-1\right) n_{2 p} & -\frac{1}{2} n_{1 p} \left(n_{2 p}-1\right) & \frac{1}{2} \left(-n_{2 p} n_{1 p}+n_{1 p}+n_{2 p}\right) \\
 \frac{1}{2} \left(n_{1 p}-1\right) n_{2 p} & \frac{1}{2} \left(n_{1 p}-2\right) n_{2 p} & \frac{1}{2} \left(n_{1 p} \left(n_{2 p}-1\right)-n_{2 p}+2\right) & \frac{1}{2} \left(n_{1 p}-2\right) \left(n_{2 p}-1\right) \\
 \frac{1}{4} n_{1 p} n_{2 p} & \frac{1}{4} n_{1 p} n_{2 p} & \frac{1}{4} n_{1 p} n_{2 p} & \frac{1}{4} n_{1 p} n_{2 p} \\
\end{array}
\right),
\end{eqnarray}
where $n_{ip}=n(\omega'_i,T_1)+n(\omega'_i,T_2)$.


When the quantum coherence terms are included, the equations of motion can be computed by the Redfield equation, which is the equation of motion of the density operator with the coherence coupling considered. The matrix elements $M_{ij}^{lk}$ for the population terms are the same as the ones above. The elements involving coherence couplings are given as follows \cite{wang2019nonequilibrium}:
\begin{align}
M_{11}^{23}&=M_{11}^{32}= - \sin(\theta/2) \cos(\theta/2)\ \big(\Gamma_1( n_1^{T_1}- n_1^{T_2})+ \Gamma_2 (n_2^{T_1}- n_2^{T_2})\big),\\
M_{22}^{23}&=M_{22}^{32}=M_{33}^{23}=M_{33}^{32}=M_{23}^{11}=M_{23}^{22}=M_{23}^{33}=M_{23}^{44}=-M_{11}^{23}, \\
M_{44}^{23}&=M_{44}^{32}=M_{11}^{23}, \\
M_{23}^{32}&=-\Gamma_1-\Gamma_2+i(\omega'_2-\omega'_1).
\end{align}
Here,  $\omega'_{1,2}=\frac{1}{2} (\omega_1+\omega_2 \pm \sqrt{(\omega_1-\omega_2)^2+4 \Delta^2})$, $M_{32}^{lk}=(M_{23}^{lk})^*$ for all $l,k$ and the rest of the matrix elements are zero. 

The eigenvalues of the transition matrix is given by
\begin{gather}
    \lambda_1=-4\Gamma,\, \lambda_2=\lambda_3=-2\Gamma,\, \lambda_{4,5}= -2\Gamma\pm 2\Delta i,\, \lambda_6=0\,. 
\end{gather}
Notably, the transition matrix has two additional complex eigenvalues $\lambda_{4}$ and $\lambda_5$ in comparison with that in the Lindbladian dynamics. The imaginary parts in the eigenvalues result in more oscillatory behaviors in the dynamics which can postpone or advance the time of the Mpemba crossing. Importantly, under the wide-band approximation in the Lindbladian formalism, the system is characterized by the population terms in the density matrix. This form of density matrix has a classical correspondence of the probability distribution and its dynamics has a classical interpretation. The conventional Lindblad equation uses the secular approximation to decouple the coherence terms from the population terms. However, the quantum coherence terms, which does not have the classical counterpart, can couple directly with the population terms in the Redfield equation and can influence the population dynamics in a nontrivial way.

The basis transformation is explicitly given as follows. For the symmetric site case, i.e. $\omega_1=\omega_2$, the transformation from the global to the local basis is given by
\begin{eqnarray}\label{local}
    \rho_{\text{local}}=U\rho U^\dg =\begin{pmatrix}\rho_{11} & 0 & 0 & 0 \\ 0& \frac{1}{2}(\rho_{22}+\rho_{33})-\mathrm{Re} (\rho_{23}) & -\frac{1}{2}(\rho_{22}-\rho_{33})+\mathrm{Im}(\rho_{23}) & 0 \\ 0& -\frac{1}{2}(\rho_{22}-\rho_{33})-\mathrm{Im} (\rho_{23}) & \frac{1}{2}(\rho_{22}+\rho_{33}) +\mathrm{Re}(\rho_{23}) &0\\0&0&0&\rho_{44} \end{pmatrix}, 
\end{eqnarray}  
where $\rho_{ij}$ in the matrix is evaluated in the energy eigenbasis. The concurrence written in the eigenbasis of the system Hamiltonian $\mathcal{H}_S$ is given by 
\begin{gather}
    \mathcal{E}(\rho)=2\max (0,\,\frac{1}{2}|\rho_{22}-\rho_{33}|-\sqrt{\rho_{11} \rho_{44}})\,.
\end{gather}

\end{widetext}

\bibliography{pnas-sample}
\newpage
\end{document}